\documentclass[aps,prd
,preprint,tightenlines,
nofootinbib,showpacs]{revtex4}
\usepackage{amssymb,latexsym}
\usepackage{amsmath,amsbsy,bbm}
\usepackage{epsfig,bm}
\usepackage{graphicx,comment}
\DeclareMathOperator{\tr}{tr}

\begin{document}
\def\a{{\alpha}}
\def\b{{\beta}}
\def\d{{\delta}}
\def\D{{\Delta}}
\def\X{{\Xi}}
\def\e{{\varepsilon}}
\def\g{{\gamma}}
\def\G{{\Gamma}}
\def\k{{\kappa}}
\def\l{{\lambda}}
\def\L{{\Lambda}}
\def\m{{\mu}}
\def\n{{\nu}}
\def\o{{\omega}}
\def\O{{\Omega}}
\def\S{{\Sigma}}
\def\s{{\sigma}}
\def\th{{\theta}}

\def\ol#1{{\overline{#1}}}

\def\Aslash{A\hskip-0.45em /}
\def\Dslash{D\hskip-0.65em /}
\def\Dtslash{\tilde{D} \hskip-0.65em /}

\def\CPT{{$\chi$PT}}
\def\QCPT{{Q$\chi$PT}}
\def\PQCPT{{PQ$\chi$PT}}
\def\tr{\text{tr}}
\def\str{\text{str}}
\def\diag{\text{diag}}
\def\order{{\mathcal O}}

\def\cF{{\mathcal F}}
\def\cS{{\mathcal S}}
\def\cC{{\mathcal C}}
\def\cB{{\mathcal B}}
\def\cT{{\mathcal T}}
\def\cQ{{\mathcal Q}}
\def\cL{{\mathcal L}}
\def\cO{{\mathcal O}}
\def\cA{{\mathcal A}}
\def\cQ{{\mathcal Q}}
\def\cR{{\mathcal R}}
\def\cH{{\mathcal H}}
\def\cW{{\mathcal W}}
\def\cM{{\mathcal M}}
\def\cD{{\mathcal D}}
\def\cN{{\mathcal N}}
\def\cP{{\mathcal P}}
\def\cK{{\mathcal K}}
\def\Qt{{\tilde{Q}}}
\def\Dt{{\tilde{D}}}
\def\St{{\tilde{\Sigma}}}
\def\cBt{{\tilde{\mathcal{B}}}}
\def\cDt{{\tilde{\mathcal{D}}}}
\def\cTt{{\tilde{\mathcal{T}}}}
\def\cMt{{\tilde{\mathcal{M}}}}
\def\At{{\tilde{A}}}
\def\cNt{{\tilde{\mathcal{N}}}}
\def\cOt{{\tilde{\mathcal{O}}}}
\def\cPt{{\tilde{\mathcal{P}}}}
\def\cI{{\mathcal{I}}}
\def\cJ{{\mathcal{J}}}

\def\eqref#1{{(\ref{#1})}}

\preprint{UMD-40762-467}

\title{Meson-Baryon Scattering Parameters from Lattice QCD with an Isospin Chemical Potential}

\author{Paulo F. Bedaque}
\email[]{bedaque@umd.edu}
\affiliation{%
Maryland Center for Fundamental Physics, 
Department of Physics, 
University of Maryland, 
College Park,  
MD 20742-4111, 
USA
}

\author{Michael I. Buchoff}
\email[]{mbuchoff@umd.edu}
\affiliation{%
Maryland Center for Fundamental Physics, 
Department of Physics, 
University of Maryland, 
College Park,  
MD 20742-4111, 
USA
}

\author{Brian C. Tiburzi}
\email[]{bctiburz@umd.edu}
\affiliation{%
Maryland Center for Fundamental Physics, 
Department of Physics, 
University of Maryland, 
College Park,  
MD 20742-4111, 
USA
}

\date{\today}


\begin{abstract}
The extraction of meson-baryon scattering parameters from lattice QCD is complicated by the necessity, in some channels, of including annihilation diagrams. We consider a strategy to avoid the need for these extremely costly and noisy contributions. The strategy is based on simulations with  an isospin chemical potential which, contrary to a baryon chemical potential, has no sign problem.
When the isospin chemical potential is larger than a critical value, 
a charged pion condensate forms.
Baryons propagating in the pion condensate will have their masses modified
by pion-baryon scattering parameters.
Consequently these parameters can be extracted from lattice QCD simulations in an isospin chemical potential, 
and we detail precisely which low-energy constants using baryon chiral perturbation theory. 
\end{abstract}
\maketitle

\section{Introduction}                                        %

In recent years, 
the determination of hadronic interactions via lattice QCD has shown substantial progress.  
In particular, 
calculations of meson-meson scattering lengths, 
such as that in the 
$I=2$ channel of pion-pion scattering, 
agree to within a few percent of experiment~%
\cite{Yamazaki:2004qb,Beane:2005rj,Beane:2007xs,Feng:2009ij,Beane:2007uh}.  Additionally, 
within the past year,   
considerable progress has been made with multi-meson systems~%
\cite{Beane:2007es,Detmold:2008fn,Detmold:2008yn}, 
two and three baryon interactions%
~\cite{Beane:2006mx,Beane:2006gf,Beane:2009gs}, 
and several meson-baryon scattering processes~%
\cite{Torok:2009dg}.      
The computation of a class of hadronic interactions, 
however, 
is hindered by the existence of annihilation diagrams, 
which, 
using current lattice methods, 
require all-to-all propagators.  
While some progress has been made in the computation of these propagators~%
\cite{Peardon:2009gh}, 
the scaling to larger lattices is still prohibitively expensive for practical calculations of hadronic interactions. 
As a result, 
information on pion-nucleon scattering, 
and 
$\ol K N$ scattering, 
a process thought to be important for kaon condensation in neutron stars~%
\cite{Kaplan:1986yq,Nelson:1987dg,Brown:1987hj,Brown:1992ib,Brown:2007ara}, 
has remained elusive to a first principles lattice calculation.

In this work, 
we propose a novel method for extracting the low-energy constants (LECs) for scattering processes with or without annihilation diagrams by employing an isospin chemical potential.  In terms of lattice calculations, 
the addition of an isospin chemical potential results in a positive definite determinant, 
which avoids the fermion sign problem that plagues other finite density calculations,  
for more information and examples, see~%
\cite{Stephanov:1996ki,Barbour:1997ej,Alford:1998sd,deForcrand:1999cy,Cox:1999nt}.  
An additional feature of an isospin chemical potential intrinsic to our analysis is the formation of a pion condensate when the chemical potential reaches a critical value~%
\cite{Son:2000xc,Son:2000by}.  
This feature has been observed in previous lattice calculations~%
\cite{Kogut:2002tm,Kogut:2002zg,Kogut:2002cm,Kogut:2003ju,Kogut:2004zg,deForcrand:2007uz}.  
In the presence of this dynamically generated pion condensate, 
propagating baryons have their masses shifted by an amount depending on the value of the condensate, 
chemical potential, 
and a linear combination of LECs, 
with a dependence that can be calculated in heavy baryon chiral perturbation theory 
(HB\CPT).   
As a result, 
precision measurement of baryon masses for several values of the chemical potential coupled with the HB\CPT\ analysis gives an avenue for extracting scattering parameters, even for processes with annihilation diagrams.

The organization of our work is as follows.  
In Sec.~\ref{s:Iso}, 
the continuum QCD action with an isospin chemical potential is detailed along with the 
leading-order values for the condensates.    
In Sec.~\ref{s:baryonsSU2}, 
we present the two-flavor HB\CPT\ analysis of nucleon and hyperon masses in the presence of the pion condensate. 
Our results show the linear combinations of LECs that can be extracted from doing several measurements at different chemical potentials above the critical point.  
In Sec.~\ref{s:Strange}, 
the results of three-flavor HB\CPT\ are derived by matching conditions to the two-flavor results.
A discussion in Sec.~\ref{s:End} concludes our work.

\section{Isospin Chemical Potential}       %
\label{s:Iso}                                              %

Let us consider QCD with two light-quark flavors.  
The strange quark will be included below in Sect.~\ref{s:Strange}. 
Using a continuum Minkowski space action, 
the matter part of the QCD Lagrange density is 
\begin{equation} \label{eq:L}
\cL 
=
\ol \psi 
\left[ 
i 
\left(
\Dslash 
+
i
\mu_I
\gamma_0
\frac{\tau^3}{2}
\right)
-
m_q
+
i 
\epsilon
\gamma_5
\frac{\tau^2}{2}
\right]
\psi
,\end{equation}
where 
$\Dslash \,$ 
is the 
$SU(3)_c$ 
gauge covariant derivative,  
$m_q = \diag (m, m)$
is the quark mass matrix in the isospin limit, 
and 
$\psi = ( u, d )^T$ an isodoublet of quarks. 
The isospin chemical potential, 
$\mu_I$,
appears as the time-component of a uniform gauge field. 
The term proportional to 
$\epsilon$
is an explicit isospin breaking term. 
It is included here because the spontaneous  breaking of isospin is an essential part of our analysis. 
Because symmetries do not break spontaneously at finite volumes, 
a small explicit breaking is necessary.  
This explicit breaking should go to zero as the size of the lattice goes to infinity.%
\footnote{
In Euclidean space, 
the action can be written in the form 
$S_E = \ol \psi ( \cD + m_q ) \psi$,
where
\begin{equation}
\cD = \Dslash + \mu_I \gamma_4 \frac{\tau^3}{2} + i \epsilon \gamma_5 \frac{\tau^2}{2} 
.\end{equation} 
The Dirac operator 
$\cD$ 
satisfies the condition
$\tau^1 \gamma_5 \, \cD \, \tau^1 \gamma_5 = \cD^\dagger$, 
which ensures that the product of $u$ and $d$ fermionic determinants is positive.
}

For vanishing quark masses, 
$m=0$,
and vanishing external sources,
$\mu_I = \epsilon = 0$, 
the QCD action has an 
$SU(2)_L \otimes SU(2)_R$
symmetry that is spontaneously 
broken down to 
$SU(2)_V$. 
For non-vanishing chemical potential, 
the QCD action has only a
$U(1)_L \otimes U(1)_R$
symmetry associated with $I_3$. 
This will be broken down to
$U(1)_V$
by both spontaneous chiral symmetry breaking 
and the explicit symmetry breaking introduced by 
the quark mass. 
At finite 
$\mu_I$ 
and finite
$\epsilon$, 
the only continuous symmetry of 
$\cL$
is the 
$U(1)_B$
associated with baryon number.
\footnote{
With both 
$\mu_I$ 
and 
$\epsilon$
non-vanishing there is no symmetry that prevents their renormalization. 
These parameters are only multiplicatively renormalized because, for example, 
the isospin chemical potential is the only term in the action that breaks 
$C \otimes \tau^3$, 
and the explicit isospin breaking term is the only term in the action that breaks parity.   
Furthermore
because we are interested in the limit of small 
$\epsilon$,
the action will be approximately symmetric under the 
$U(1)_V$ 
associated with $\tau^3$. 
Consequently 
$\mu_I$ 
cannot be appreciably renormalized. 
}
We can write down an effective field theory of QCD
that takes into account this pattern of spontaneous and explicit symmetry 
breaking. 
This effective theory is \CPT, 
which is written in terms of the coset 
$U \in SU(2)_L \otimes SU(2)_R / SU(2)_V$,
and at leading order has the form
\begin{equation} \label{eq:CPT}
\cL
=
\frac{f^2}{8}
\left[
<
D_\mu U D^\mu U^\dagger
>
+
2 \lambda 
<
M^\dagger U + U^\dagger M
>
\right]
.\end{equation}
Here the angled brackets denote flavor traces, 
and the external sources have been included in the terms
\begin{eqnarray}
M &=& s + i p ,  
\notag
\\
D_\mu U 
&=&
\partial_\mu U
+ i [ \mathbb{V}_\mu, U]
,\end{eqnarray}
where the external vector potential is given by: 
$\mathbb{V}_\mu = \mu_I \frac{\tau^3}{2} \delta_{\mu,0}$,  
the pseusdoscalar source is given by:
$p = \epsilon \frac{\tau^2}{2}$,
and the scalar source is just the quark mass,
$s = m_q$.

The isospin chemical potential favors the presence of up quarks as opposed to down quarks. When it is larger than about 
$m_\pi$, 
it becomes energetically favorable for the ground state to contain positive pions. 
Indeed, 
assuming a uniform value $U_0$ 
for the vacuum expectation value of $U(x)$, 
we can minimize the effective potential to determine
the vacuum (ground) state. 
Using the standard parametrization of 
$SU(2)$, 
namely
$U_0 = \exp [ i \alpha \, \bm{n} \cdot \bm{\tau} ]$, 
with $\bm{n} \cdot \bm{n} = 1$,
we find
\begin{equation}
U_0
=
\begin{cases}
1, & | \mu_I | < m_\pi, 
\\
\exp [ i \alpha \, \tau^2 ], &
|\mu_I| > m_\pi
\end{cases}
.\end{equation}
The angle 
$\a$
is determined, at lowest order in the chiral expansion, by the transcendental equation
\begin{equation} \label{eq:alpha}
\cos \alpha 
= 
\frac{m_\pi^2}{\mu_I^2}
- 
\frac{\lambda \epsilon}{\mu_I^2} \cot \a
,\end{equation}
where the pion mass satisfies the Gell-Mann--Oakes--Renner relation,
$m_\pi^2 = 2 \lambda m$.
For small values of 
$\epsilon$,
one can determine 
$\a$
using a perturbative expansion of Eq.~\eqref{eq:alpha}.
This expansion, however, breaks down as one nears 
the critical value of the chemical potential from above. 
In the condensed phase, 
the condensates have values
\begin{eqnarray}
\langle \ol \psi \psi \rangle 
&=&
f^2 \lambda \cos \a,
\notag
\\
i \langle \ol \psi \tau^2 \gamma_5 \psi \rangle
&=&
f^2 \lambda \sin \a
.\end{eqnarray}

\section{Baryons in an isospin chemical potential: $SU(2)$}  %
\label{s:baryonsSU2}                                                               %

When a hadron propagates in the background of a pion condensate its properties are changed due to the interactions of the hadron with the pions in the condensate. Their mass, for instance, is shifted by an amount closely related to the forward scattering amplitude of pions on the hadron. Unfortunately, the pions in the condensate are not exactly on-shell and no model independent relation can be found between pion-hadron scattering amplitudes and the mass shift. However, the chiral expansion of these two quantities are be related in the way described in this section.

\subsection{Nucleons}      %

We can address effects of the isospin chemical potential 
on nucleons by using the heavy nucleon chiral Lagrangian~\cite{Bernard:1992qa}. 
This effective theory accounts for spontaneous and explicit 
chiral symmetry breaking in the nucleon sector. The theory is 
written in terms of static nucleon fields, and thereby the 
nucleon mass can be treated commensurately with the chiral symmetry
breaking scale.

The leading-order heavy nucleon chiral Lagrangian is written 
in terms of the nucleon doublet,
$N = (p, n)^T$,
and has the form
\begin{eqnarray} \label{eq:nuke}
\cL^{(1)}
&=&
N^\dagger \left( i v \cdot D + 2 g_A S \cdot A \right) N
.\end{eqnarray}
The vector $v_\mu$ is the nucleon four-velocity, 
while the vector, 
$V_\mu$,  
and axial-vector,
$A_\mu$,
fields of pions
are defined by 
\begin{eqnarray}
V_\mu 
&=&
\frac{1}{2} 
\left(  
\xi^\dagger_L D^L_\mu \xi_L
+
\xi_R D^R_\mu \xi_R^\dagger
\right)
\notag
,\\
A_\mu
&=& 
\frac{i}{2} 
\left(  
\xi^\dagger_L D^L_\mu \xi_L
-
\xi_R D^R_\mu \xi_R^\dagger
\right)
.\end{eqnarray}
The former appears in the chirally covariant derivative
$D_\mu$, 
which has the following action on the nucleon field
\begin{equation}
(D_\mu N)_i
=
\partial_\mu N_i
+ 
(V_\mu)_{i} {}^j N_j
.\end{equation} 
Appearing in the vector and axial-vector pion fields are 
$\xi_L$, 
and 
$\xi_R$  
which arise from the definition
$U = \xi_L \xi_R$. 
Under a chiral transformation,
$(L,R) \in SU(2)_L \otimes SU(2)_R$, 
the pion fields transform as:
$\xi_L(x) \to L \, \xi_L(x) \mathcal{U}(x)$,
and
$\xi_R(x) \to \mathcal{U}^\dagger(x) \xi_R(x) R^\dagger$, 
where $\mathcal{U}(x)$ is the matrix entering the transformation of the nucleon field, 
$N_i \to \mathcal{U}(x)_i {}^j N_j$. 
The left-handed and right-handed derivatives act according to the rules
\begin{eqnarray}
D^L_\mu \xi_L
&=&
\partial_\mu \xi_L
+ i \mathbb{L}_\mu \xi_L
\notag
,\\
D^R_\mu \xi_R
&=&
\partial_\mu \xi_R
-
i \xi_R \mathbb{R}_\mu 
.\end{eqnarray}
For the case of an external vector field, 
the left- and right-handed vector sources coincide:
$\mathbb{L}_\mu = \mathbb{R}_\mu = \mathbb{V}_\mu$. 
In an isospin chemical potential, 
the spurions are given the final values
\begin{eqnarray}
\xi_L(x) &=& \xi_0 \,  \xi(x),
\notag
\\
\xi_R(x) &=& \xi(x) \xi_0,
\end{eqnarray}
so that the vacuum value is 
$U_0 = \xi_0^2$.

At  leading-order using Lagrangian Eq.~\eqref{eq:nuke}, 
the shift in nucleon mass due to the chemical potential is the trivial result
\begin{equation}
M_N 
= 
M_{N}^{(0)}  -  \mu_I \cos \a \frac{\tau^3}{2}
,\end{equation}
where 
$M_N^{(0)}$ 
is the chiral limit value. 
Beyond leading-order, 
the nucleon mass receives corrections that depend on low-energy constants. 
These constants are the coefficients of terms in the second-order nucleon chiral Lagnrangian. 
Including relativistic corrections with fixed coefficients, 
the complete
\footnote{
When one considers external sources with non-vanishing field strength tensors, 
additional terms are present. 
The external sources we consider, 
however, 
are uniform in spacetime. 
} 
 second-order Lagrangian is 
\begin{eqnarray} 
\cL^{(2)}
&=&
N^\dagger 
\left[
- 
\frac{D_\perp^2}{2 M_{N}^{(0)}} 
+
\frac{1}{2 M}
[S^\mu, S^\nu] \, [ D_\mu, D_\nu]
- 
\frac{i g_A}{M_N^{(0)}}
\Big\{
v \cdot A, S \cdot D
\Big\}
+
4 
c_1 
< \cM_+ >
\right.
\notag \\
&& \phantom{s}
\left.
+
4 
\left(c_2
- 
\frac{g_A^2}{8 M_N^{(0)}}
\right)
(v \cdot A)^2
+
4 c_3 A^2
+ 
4 c_4 [S^\mu, S^\nu] A_\mu A_\nu
+
4 c_5 \tilde{\cM}_+
\right] N,
\label{eq:secondnuke}
\end{eqnarray}
where 
$\tilde{\cM}_+ = \cM_+ - \frac{1}{2}  < \cM_+ >$, 
and 
$\cM_+ = \frac{1}{2} \lambda (\xi^\dagger_L M \xi^\dagger_R + \xi_R M^\dagger \xi_L)$. 
The last term with coefficient $c_5$ only contributes in the presence of strong isospin breaking.

Utilizing the second-order nucleon Lagrangian, 
we arrive at the nucleon mass
\begin{equation}
M_N 
=
M_N^{(0)} 
- 
\mu_I \cos \a \frac{\tau^3}{2}
+
4 c_1 \left( m_\pi^2 \cos \a + \lambda \epsilon \sin \a \right)
+
\left( c_2 - \frac{g_A^2}{8 M}  + c_3 \right) \mu_I^2 \sin^2 \a
\end{equation}
The second-order correction is entirely isoscalar,
and allows access to the low-energy constant
$c_1$, 
and 
the combination 
$c_2 + c_3$. 
One can vary the quark mass and isospin chemical potential 
to isolate these coefficients from the observed behavior of the nucleon mass in lattice QCD.
\footnote{
One can go further and impose isospin twisted boundary conditions 
on the quark fields, 
e.g.~$\psi(x + L \hat{z}) = \exp [ i \theta \tau^3 ] \psi(x)$.
This has the advantage of introducing non-vanishing spatial components of the vacuum value of 
$A_\mu$,
and leads to the ability to isolate more low-energy constants.
For the example mentioned, 
the 
$c_3$ 
coefficient can be determined from the mixing angle between nucleons. 
}

\subsection{Hyperons}                         %
\label{s:hyper}                                      %

Recently there has been interest in treating hyperons using 
$SU(2)$ \CPT~\cite{Beane:2003yx,Frink:2002ht,Tiburzi:2008bk,Jiang:2009sf,Mai:2009ce,Tiburzi:2009cf}.
We consider hyperons in an isospin chemical potential, 
and determine which low-energy constants of $SU(2)$ 
\CPT\ 
can be determined from their masses.
This is a natural approach because the isospin chemical potential 
produces only pion-hyperon interactions.

In the stangeness $S=1$ sector are the $\L$ and $\S$ baryons. 
The 
$\L$ 
baryon is an isosinglet, 
while the 
$\S$ 
baryons form an isotriplet. 
The leading-order Lagrangian in the $S=1$ sector has the form
\begin{equation}
\cL^{(1)}
=
\Lambda^\dagger i v \cdot \partial \Lambda 
+
< 
\Sigma^\dagger \left( i v \cdot D - \D_{\L\S}  \right) \Sigma 
>
.\end{equation}
Here we have packaged the $\S$ fields into an adjoint matrix
\begin{equation}
\S = 
\begin{pmatrix}
\frac{1}{\sqrt{2}} \Sigma^0 & \Sigma^+ \\
\Sigma^- & - \frac{1}{\sqrt{2}} \Sigma^0
\end{pmatrix}
,\end{equation}
and use angled brackets to denote flavor traces. 
The covariant derivative acts as 
$D_\mu \Sigma = \partial_\mu \S + [ V_\mu, \S]$. 
The parameter 
$\D_{\L\S}$ 
is the 
$SU(2)$ 
chiral limit value of the mass splitting,
$\D_{\L\S} = M^{(0)}_\S - M^{(0)}_\L$. 
In writing the leading-order Lagrangian, 
we have omitted the axial couplings to pions 
as these terms are not relevant for our computation. 
Similarly
we write down only the contributing terms from the 
second-order 
$S=1$ 
Lagrangian
\footnote{
To account for Kroll-Ruderman terms in the $S=1$ sector, 
replace
$c_2^\L \to c_2^\L - \frac{1}{6} g_{\L \S}^2 / (M^{(0)}_\L + M^{(0)}_\S)$,
$c_2^\S \to c_2^\S  - \frac{1}{4} g_{\S\S}^2 / (M^{(0)}_\L + M^{(0)}_\S)$,
and  
$c_6^\S \to c_6^\S - \frac{1}{12}  g_{\L \S}^2 / M^{(0)}_\L + \frac{1}{8} g_{\S\S}^2 / (M^{(0)}_\L + M^{(0)}_\S)$.
The normalization of the hyperon axial couplings 
$g_{\S\S}$ 
and 
$g_{\L \S}$ 
is that of~\cite{Tiburzi:2008bk}. 
}
\begin{eqnarray}
\cL 
&=&
\L^\dagger \L  
\left[
4 c_1^\L
 < \cM_+ >
+
2 c_2^\L
< v \cdot A^2  >
+ 
2 c_3^\L
<  A^2 >
\right]
\notag \\
&&
+ 
< 
\Sigma^\dagger \Sigma
>
\left[
4 c_1^\S
< \cM_+ >
+
2
c_2^\S
< v \cdot A^2 >
+
2
c_3^\S
< A^2 >
\right]
\notag \\
&& +
4 c_6
< \S^\dagger  v \cdot A> <  v \cdot A \, \S >
+
4 c_7 
< \S^\dagger A_\mu > < A^\mu \S >
.\end{eqnarray}

Using the first- and second-order Lagrangians, 
we find the mass of the 
$\L$ 
baryon
\begin{equation}
M_\L 
= 
M_\L^{(0)}
+
4 c_1^\L 
\left( 
m_\pi^2 \cos \a 
+ 
\lambda \epsilon \sin \a 
\right)
+
(c_2^\L + c_3^\L)
\mu_I^2 \sin^2 \a
,\end{equation}
and see that the low-energy constant 
$c_1^\L$, 
and the combination 
$c_2^\L + c_3^\L$
can be accessed. 
For the 
$\S^0$ 
baryon,
we find the mass
\begin{eqnarray}
M_{\S^0}
&=&
M_{\S}^{(0)}
+
4 c_1^\S 
\left( 
m_\pi^2 \cos \a 
+ 
\lambda \epsilon \sin \a 
\right)
+
(c_2^\S + c_3^\S) 
\mu_I^2 \sin^2 \a
.\end{eqnarray}
At this order in the chiral expansion, 
the charged 
$\S$ 
baryons mix. 
We find the mass eigenvalues
\begin{eqnarray}
M_\pm
&=&
M_{\S^0}
+
( c_6^\S + c_7^\S)
\mu_I^2 \sin^2 \a
\pm 
\mu_I
\sqrt{
\cos^2 \a
+
(c_6^\S + c_7^\S)^2 
\mu_I^2 \sin^4 \a
}
.\end{eqnarray}
The corresponding eigenstates are
\begin{eqnarray}
| + \rangle
&=& 
\cos \frac{\Theta}{2} 
| \S^+ \rangle
-
\sin \frac{\Theta}{2} \,
| \S^- \rangle
\notag
\\
| - \rangle
&=&
\sin \frac{\Theta}{2} \,
| \S^+ \rangle
-
\cos \frac{\Theta}{2} \,
| \S^- \rangle 
,\end{eqnarray}
where the mixing angle can be found from
\begin{equation}
\cos \Theta
= 
- \frac{\cos \a}
{\sqrt{ 
\cos^2 \a
+
(c_6^\S + c_7^\S)^2 
\mu_I^2 \sin^4 \a
}}
.\end{equation}
Finally because the 
$c_j^\S$ 
coefficients should scale with the inverse of the chiral symmetry breaking scale, 
the mixing between 
$\S$ 
baryons should be small for moderately sized 
$\mu_I$.

In the strangeness $S=2$ sector are the cascades. 
The case of the cascade is similar to the nucleon
as the cascades transform as an isospin doublet, 
$\Xi = ( \Xi^0, \Xi^- )^T$. 
Consequently the first- and second-order
Lagrangians have the same form as those for the 
nucleon, Eqs.~\eqref{eq:nuke} and \eqref{eq:secondnuke}.
One merely replaces the chiral limit mass of the nucleon, 
$M_N^{(0)}$,
with the chiral limit mass of the cascade, 
$M_{\X}^{(0)}$, 
the nucleon axial charge, 
$g_A$
with the cascade axial charge, 
$g_{\X\X}$. 
Finally each of the low-energy constants
$c_j$ 
have different values for the cascades;
these we denote by $c^{\X}_j$.
The cascade masses are thus given by
\begin{equation}
M_\X
=
M^{(0)}_\X 
- 
\mu_I \cos \a \frac{\tau^3}{2}
+
4 c^\X_1 \left( m_\pi^2 \cos \a + \lambda \epsilon \sin \a \right)
+
\left( c^\X_2 - \frac{g_{\X\X}^2}{8 M_\X^{(0)}}  + c^\X_3 \right) \mu_I^2 \sin^2 \a
.\end{equation}
Varying the quark mass and isospin chemical potential allows one 
to access 
$c_1^\X$, 
and the combination 
$c_2^\X + c_3^\X$.

\section{Strange Quark and $SU(3)$}                              %
\label{s:Strange}                                                               %

We can repeat the above analysis including a third light quark.
The isospin chemical potential is added proportional to the 
$\l^3$ Gell-Mann matrix, 
while the explicit isospin breaking term we choose in the direction of 
$\l^2$. 
For a massless strange quark, 
the QCD action with 
$\mu_I$ 
and 
$\epsilon$ 
both non-vanishing possesses a 
$U(1)_L \otimes U(1)_R$ 
symmetry associated with strangeness. 
The strange quark mass breaks this symmetry 
down to 
$U(1)_V$. 
This symmetry precludes tilting the vacuum value of 
$U$
in any strange direction. 
In 
$SU(3)$,  
we must have
\begin{equation}
U_0 
= 
\begin{pmatrix}
e^{ i \a \tau^2 } & 0 \\
0 & 1
\end{pmatrix} 
.\end{equation}
Consequently the non-vanishing values of the condensates are 
\begin{eqnarray}
\frac{1}{3} \langle \ol \psi (1 - \sqrt{3} \l^8) \psi \rangle
&=&
\lambda f^2,
\notag
\\
\frac{1}{3} \langle \ol \psi (2 + \sqrt{3} \l^8 ) \psi \rangle 
&=& 
\lambda f^2 \cos \a
\notag
,\\
i \langle \ol \psi \l^2 \gamma_5 \psi \rangle 
&=& 
\l f^2 \sin \a
.\end{eqnarray}

It is now straightforward to determine the effects of the isospin chemical potential on the octet baryons. 
The octet baryons can be packaged into an $SU(3)$ adjoint matrix 
$B$, 
given by
\begin{equation}
B
= 
\begin{pmatrix}
\frac{1}{\sqrt{2}} \S^0 + \frac{1}{\sqrt{6}} \L & \S^+ & p \\
\S^- & - \frac{1}{\sqrt{2}} \S^0 + \frac{1}{\sqrt{6}} \L & n \\
\X^- & \X^0 & - \frac{2}{\sqrt{6}} \L
\end{pmatrix}
.\end{equation}
The relevant term of the leading-order 
$SU(3)$
Lagrangian is
\begin{equation}
\cL^{(1)} 
=
< B^\dagger [ i v \cdot D, B] > 
,\end{equation}
with 
$D_\mu B = \partial_\mu B + [ V_\mu, B]$. 
The kinetic term reproduces the 
isospin-dependent mass shifts 
for the octet baryons found above.

The contributing terms of the second-order 
$SU(3)$ 
chiral Lagrangian are
\begin{eqnarray}
\cL^{(2)}
&=&
2 b_F
< B^\dagger [\cM_+ , B] >
+
2 b_D
< B^\dagger \{ \cM_+, B \} >
+ 
2 b_0
<B^\dagger B> < \cM_+>
\notag \\
&&+
b_1 
< B^\dagger [ A_\mu, [ A^\mu, B] ] >
+
b_2
< B^\dagger [ A_\mu, \{ A^\mu, B \} ] >
+ 
b_3
< B^\dagger \{ A_\mu, \{ A^\mu, B \} \} >
\notag \\
&&+
b_4
< B^\dagger [ v \cdot A, [ v \cdot A, B ] ] > 
+ 
b_5
< B^\dagger [ v \cdot A, \{ v \cdot A, B \} ] >
\notag \\
&&+
b_6
< B^\dagger \{ v \cdot A, \{ v \cdot A, B \} \} >
+ 
b_7
< B^\dagger B > < A^2 > 
+ 
b_8 
< B^\dagger B> < v \cdot A^2 >
.
\notag \\
\end{eqnarray}
Evaluating the masses of the octet baryons, 
we arrive at results identical to the 
$SU(2)$
case but with  
$SU(3)$ 
relations between parameters.
The matching of parameters for the baryon masses 
have been given in~\cite{Tiburzi:2008bk}, 
while those for pion-baryon 
scattering have been given in~\cite{Mai:2009ce}. 
Terms proportional to the strange quark mass
give rise to shifts of the 
$SU(3)$ chiral limit value of the baryon mass.
Such terms are the leading contributions in 
$SU(3)$
to the 
$SU(2)$ 
chiral limit values of the baryon masses employed above. 
As our interest is with the 
$\mu_I$ 
dependence, 
we do not duplicate the matching relations between 
the chiral limit baryon masses here.
Matching of 
$\cM_+$ 
operators in the upper 
$2 \times 2$ 
block yields the relations~\cite{Tiburzi:2008bk}
\begin{eqnarray}
4 c_1 &=& 
b_D + b_F + 2 b_0,
\notag
\\
4 c_1^\L &=& 
\frac{2}{3} b_D + 2 b_0,
\notag
\\
4 c_1^\S &=&
2  b_D + 2 b_0,
\notag
\\
4 c_1^\X &=&
b_D - b_F + 2 b_0, 
.\end{eqnarray}
Matching of operators involving the product of two axial-vector pion fields gives the relations~\cite{Mai:2009ce}
\begin{eqnarray}
4(c_2 + c_3)
&=&
(b_1 + b_4) +  (b_2  + b_5) + ( b_3 + b_6)  + 2 ( b_7 + b_8)
\notag
,\\
4 (c_2^\L + c_3^\L)
&=&
\frac{4}{3} ( b_3 +  b_6 ) +   2 ( b_7 + b_8)
\notag
,\\
4( c_2^\S + c_3^\S )
&=&
4 ( b_1 + b_4)  + 2 ( b_7 + b_8 )
\notag
,\\
4( c_6^\S + c_7^\S )
&=&
- 2 ( b_1 + b_4)  + 2  ( b_3  + b_6)
\notag
,\\
4( c_2^\X + c_3^\X )
&=&
( b_1  + b_4) -  ( b_2 + b_5)  +  (b_3 + b_6)  + 2 (b_7 + b_8)
.\end{eqnarray}
We have subsumed the Kroll-Ruderman terms here. 
From these matching conditions, 
we see that all three of the LECs for the quark mass dependent
operators can be determined, 
along with four combinations 
of the eight double axial-insertion operators. 
The overdetermination of these parameters, 
moreover, 
serves as a useful check on 
$SU(3)$ 
symmetry.

\section{Discussion}                                                         %
\label{s:End}                                                                    %

Above, 
we show what aspects of meson-baryon scattering can be determined
from simulating lattice QCD with an isospin chemical potential.
Our approach allows for the determination of the relevant LECs from first principles; 
and,
specifically for pion-nucleon processes,
avoids the need to evaluate all-to-all propagators. 
Several attempts at extracting  the non-derivative pion-nucleon couplings from analysis of experimental data have been attempted in the past. 
In \cite{Rentmeester:2003mf}, 
nucleon-nucleon scattering data was used to determine the pion-nucleon constants through their effect on two-pion exchange contributions to the nuclear force. Pion-nucleon scattering data has also been used for this purpose, for instance \cite{Fettes:1998ud}. 
Significant uncertainties and discrepancies between different extractions remain. 
Our method can help resolve this situation.   
The values of the antikaon-nucleon couplings are considerably more uncertain, 
although attempts of extracting them from experimental data have been made, 
see, for example,~\cite{Oller:2000fj,Meissner:2004jr,Borasoy:2005ie,Oller:2005ig}.
Our framework can shed light on this issue by testing 
$SU(3)$ 
symmetry predictions.

Of particular interest is the coefficient 
$c_1$, 
related to the nucleon 
$\sigma$-term, 
which is important for addressing dark matter interactions with nuclei, 
and has been the subject of numerous studies.
The value estimated in \cite{Gasser:1990ce} has been challenged~
\cite{Buettiker:1999ap}
. 
The pion-nucleon constants also affect the dependence of the nucleon mass on the quark mass at fourth order in the chiral expansion,  
and this dependence has been used to extract them from lattice QCD calculations at different quark masses (for a recent extraction, see \cite{Ohki:2009mt}). 
These extractions are complicated, however, by the fact that the quark mass dependence of nucleon mass is barely compatible with chiral perturbation theory predictions at the current values of quark masses \cite{WalkerLoud:2008bp,WalkerLoud:2008pj}. 
The method proposed here adds another tool that is somewhat complementary to the quark mass variation of the nucleon mass. In fact, in order to extract the pion-nucleon couplings one needs to know the nucleon mass for several values of the quark mass small enough so that chiral perturbation theory is trustable. But, apparently, this range might be small \cite{WalkerLoud:2008pj}. The dependence of the nucleon mass on the isospin chemical potential adds another handle that can be used for this purpose. The computational costs of changing the quark mass or the chemical potential are comparable.  Of course, gauge configurations at different quark masses are also useful for other purposes besides nucleon mass extraction. The dependence on $\mu_I$, moreover, can also help in the extraction of other observables, making gauge configurations with $\mu_I$ more generally useful, and our proposal a realistic option.

\begin{acknowledgments}
This work is supported by the 
U.S.~Dept.~of Energy,
Grant No.~DE-FG02-93ER-40762.
\end{acknowledgments}

\appendix %

\bibliography{hb} %

\begin{thebibliography}{49}
\expandafter\ifx\csname natexlab\endcsname\relax\def\natexlab#1{#1}\fi
\expandafter\ifx\csname bibnamefont\endcsname\relax
  \def\bibnamefont#1{#1}\fi
\expandafter\ifx\csname bibfnamefont\endcsname\relax
  \def\bibfnamefont#1{#1}\fi
\expandafter\ifx\csname citenamefont\endcsname\relax
  \def\citenamefont#1{#1}\fi
\expandafter\ifx\csname url\endcsname\relax
  \def\url#1{\texttt{#1}}\fi
\expandafter\ifx\csname urlprefix\endcsname\relax\def\urlprefix{URL }\fi
\providecommand{\bibinfo}[2]{#2}
\providecommand{\eprint}[2][]{\url{#2}}

\bibitem[{\citenamefont{Yamazaki et~al.}(2004)}]{Yamazaki:2004qb}
\bibinfo{author}{\bibfnamefont{T.}~\bibnamefont{Yamazaki}} \bibnamefont{et~al.}
  (\bibinfo{collaboration}{CP-PACS}), \bibinfo{journal}{Phys. Rev.}
  \textbf{\bibinfo{volume}{D70}}, \bibinfo{pages}{074513}
  (\bibinfo{year}{2004}), \eprint{hep-lat/0402025}.

\bibitem[{\citenamefont{Beane et~al.}(2006{\natexlab{a}})\citenamefont{Beane,
  Bedaque, Orginos, and Savage}}]{Beane:2005rj}
\bibinfo{author}{\bibfnamefont{S.~R.} \bibnamefont{Beane}},
  \bibinfo{author}{\bibfnamefont{P.~F.} \bibnamefont{Bedaque}},
  \bibinfo{author}{\bibfnamefont{K.}~\bibnamefont{Orginos}}, \bibnamefont{and}
  \bibinfo{author}{\bibfnamefont{M.~J.} \bibnamefont{Savage}}
  (\bibinfo{collaboration}{NPLQCD}), \bibinfo{journal}{Phys. Rev.}
  \textbf{\bibinfo{volume}{D73}}, \bibinfo{pages}{054503}
  (\bibinfo{year}{2006}{\natexlab{a}}), \eprint{hep-lat/0506013}.

\bibitem[{\citenamefont{Beane et~al.}(2008{\natexlab{a}})}]{Beane:2007xs}
\bibinfo{author}{\bibfnamefont{S.~R.} \bibnamefont{Beane}}
  \bibnamefont{et~al.}, \bibinfo{journal}{Phys. Rev.}
  \textbf{\bibinfo{volume}{D77}}, \bibinfo{pages}{014505}
  (\bibinfo{year}{2008}{\natexlab{a}}), \eprint{0706.3026}.

\bibitem[{\citenamefont{Feng et~al.}(2009)\citenamefont{Feng, Jansen, and
  Renner}}]{Feng:2009ij}
\bibinfo{author}{\bibfnamefont{X.}~\bibnamefont{Feng}},
  \bibinfo{author}{\bibfnamefont{K.}~\bibnamefont{Jansen}}, \bibnamefont{and}
  \bibinfo{author}{\bibfnamefont{D.~B.} \bibnamefont{Renner}}
  (\bibinfo{year}{2009}), \eprint{0909.3255}.

\bibitem[{\citenamefont{Beane et~al.}(2008{\natexlab{b}})}]{Beane:2007uh}
\bibinfo{author}{\bibfnamefont{S.~R.} \bibnamefont{Beane}} \bibnamefont{et~al.}
  (\bibinfo{collaboration}{NPLQCD}), \bibinfo{journal}{Phys. Rev.}
  \textbf{\bibinfo{volume}{D77}}, \bibinfo{pages}{094507}
  (\bibinfo{year}{2008}{\natexlab{b}}), \eprint{0709.1169}.

\bibitem[{\citenamefont{Beane et~al.}(2008{\natexlab{c}})}]{Beane:2007es}
\bibinfo{author}{\bibfnamefont{S.~R.} \bibnamefont{Beane}}
  \bibnamefont{et~al.}, \bibinfo{journal}{Phys. Rev. Lett.}
  \textbf{\bibinfo{volume}{100}}, \bibinfo{pages}{082004}
  (\bibinfo{year}{2008}{\natexlab{c}}), \eprint{0710.1827}.

\bibitem[{\citenamefont{Detmold et~al.}(2008{\natexlab{a}})}]{Detmold:2008fn}
\bibinfo{author}{\bibfnamefont{W.}~\bibnamefont{Detmold}} \bibnamefont{et~al.},
  \bibinfo{journal}{Phys. Rev.} \textbf{\bibinfo{volume}{D78}},
  \bibinfo{pages}{014507} (\bibinfo{year}{2008}{\natexlab{a}}),
  \eprint{0803.2728}.

\bibitem[{\citenamefont{Detmold
  et~al.}(2008{\natexlab{b}})\citenamefont{Detmold, Orginos, Savage, and
  Walker-Loud}}]{Detmold:2008yn}
\bibinfo{author}{\bibfnamefont{W.}~\bibnamefont{Detmold}},
  \bibinfo{author}{\bibfnamefont{K.}~\bibnamefont{Orginos}},
  \bibinfo{author}{\bibfnamefont{M.~J.} \bibnamefont{Savage}},
  \bibnamefont{and}
  \bibinfo{author}{\bibfnamefont{A.}~\bibnamefont{Walker-Loud}},
  \bibinfo{journal}{Phys. Rev.} \textbf{\bibinfo{volume}{D78}},
  \bibinfo{pages}{054514} (\bibinfo{year}{2008}{\natexlab{b}}),
  \eprint{0807.1856}.

\bibitem[{\citenamefont{Beane et~al.}(2006{\natexlab{b}})\citenamefont{Beane,
  Bedaque, Orginos, and Savage}}]{Beane:2006mx}
\bibinfo{author}{\bibfnamefont{S.~R.} \bibnamefont{Beane}},
  \bibinfo{author}{\bibfnamefont{P.~F.} \bibnamefont{Bedaque}},
  \bibinfo{author}{\bibfnamefont{K.}~\bibnamefont{Orginos}}, \bibnamefont{and}
  \bibinfo{author}{\bibfnamefont{M.~J.} \bibnamefont{Savage}},
  \bibinfo{journal}{Phys. Rev. Lett.} \textbf{\bibinfo{volume}{97}},
  \bibinfo{pages}{012001} (\bibinfo{year}{2006}{\natexlab{b}}),
  \eprint{hep-lat/0602010}.

\bibitem[{\citenamefont{Beane et~al.}(2007)}]{Beane:2006gf}
\bibinfo{author}{\bibfnamefont{S.~R.} \bibnamefont{Beane}} \bibnamefont{et~al.}
  (\bibinfo{collaboration}{NPLQCD}), \bibinfo{journal}{Nucl. Phys.}
  \textbf{\bibinfo{volume}{A794}}, \bibinfo{pages}{62} (\bibinfo{year}{2007}),
  \eprint{hep-lat/0612026}.

\bibitem[{\citenamefont{Beane et~al.}(2009)}]{Beane:2009gs}
\bibinfo{author}{\bibfnamefont{S.~R.} \bibnamefont{Beane}} \bibnamefont{et~al.}
  (\bibinfo{year}{2009}), \eprint{0905.0466}.

\bibitem[{\citenamefont{Torok et~al.}(2009)}]{Torok:2009dg}
\bibinfo{author}{\bibfnamefont{A.}~\bibnamefont{Torok}} \bibnamefont{et~al.}
  (\bibinfo{year}{2009}), \eprint{0907.1913}.

\bibitem[{\citenamefont{Peardon et~al.}(2009)}]{Peardon:2009gh}
\bibinfo{author}{\bibfnamefont{M.}~\bibnamefont{Peardon}} \bibnamefont{et~al.}
  (\bibinfo{collaboration}{Hadron Spectrum}) (\bibinfo{year}{2009}),
  \eprint{0905.2160}.

\bibitem[{\citenamefont{Kaplan and Nelson}(1986)}]{Kaplan:1986yq}
\bibinfo{author}{\bibfnamefont{D.~B.} \bibnamefont{Kaplan}} \bibnamefont{and}
  \bibinfo{author}{\bibfnamefont{A.~E.} \bibnamefont{Nelson}},
  \bibinfo{journal}{Phys. Lett.} \textbf{\bibinfo{volume}{B175}},
  \bibinfo{pages}{57} (\bibinfo{year}{1986}).

\bibitem[{\citenamefont{Nelson and Kaplan}(1987)}]{Nelson:1987dg}
\bibinfo{author}{\bibfnamefont{A.~E.} \bibnamefont{Nelson}} \bibnamefont{and}
  \bibinfo{author}{\bibfnamefont{D.~B.} \bibnamefont{Kaplan}},
  \bibinfo{journal}{Phys. Lett.} \textbf{\bibinfo{volume}{B192}},
  \bibinfo{pages}{193} (\bibinfo{year}{1987}).

\bibitem[{\citenamefont{Brown et~al.}(1987)\citenamefont{Brown, Kubodera, and
  Rho}}]{Brown:1987hj}
\bibinfo{author}{\bibfnamefont{G.~E.} \bibnamefont{Brown}},
  \bibinfo{author}{\bibfnamefont{K.}~\bibnamefont{Kubodera}}, \bibnamefont{and}
  \bibinfo{author}{\bibfnamefont{M.}~\bibnamefont{Rho}},
  \bibinfo{journal}{Phys. Lett.} \textbf{\bibinfo{volume}{B192}},
  \bibinfo{pages}{273} (\bibinfo{year}{1987}).

\bibitem[{\citenamefont{Brown et~al.}(1992)\citenamefont{Brown, Thorsson,
  Kubodera, and Rho}}]{Brown:1992ib}
\bibinfo{author}{\bibfnamefont{G.~E.} \bibnamefont{Brown}},
  \bibinfo{author}{\bibfnamefont{V.}~\bibnamefont{Thorsson}},
  \bibinfo{author}{\bibfnamefont{K.}~\bibnamefont{Kubodera}}, \bibnamefont{and}
  \bibinfo{author}{\bibfnamefont{M.}~\bibnamefont{Rho}},
  \bibinfo{journal}{Phys. Lett.} \textbf{\bibinfo{volume}{B291}},
  \bibinfo{pages}{355} (\bibinfo{year}{1992}).

\bibitem[{\citenamefont{Brown et~al.}(2008)\citenamefont{Brown, Lee, and
  Rho}}]{Brown:2007ara}
\bibinfo{author}{\bibfnamefont{G.~E.} \bibnamefont{Brown}},
  \bibinfo{author}{\bibfnamefont{C.-H.} \bibnamefont{Lee}}, \bibnamefont{and}
  \bibinfo{author}{\bibfnamefont{M.}~\bibnamefont{Rho}},
  \bibinfo{journal}{Phys. Rept.} \textbf{\bibinfo{volume}{462}},
  \bibinfo{pages}{1} (\bibinfo{year}{2008}), \eprint{0708.3137}.

\bibitem[{\citenamefont{Stephanov}(1996)}]{Stephanov:1996ki}
\bibinfo{author}{\bibfnamefont{M.~A.} \bibnamefont{Stephanov}},
  \bibinfo{journal}{Phys. Rev. Lett.} \textbf{\bibinfo{volume}{76}},
  \bibinfo{pages}{4472} (\bibinfo{year}{1996}), \eprint{hep-lat/9604003}.

\bibitem[{\citenamefont{Barbour et~al.}(1998)\citenamefont{Barbour, Morrison,
  Klepfish, Kogut, and Lombardo}}]{Barbour:1997ej}
\bibinfo{author}{\bibfnamefont{I.~M.} \bibnamefont{Barbour}},
  \bibinfo{author}{\bibfnamefont{S.~E.} \bibnamefont{Morrison}},
  \bibinfo{author}{\bibfnamefont{E.~G.} \bibnamefont{Klepfish}},
  \bibinfo{author}{\bibfnamefont{J.~B.} \bibnamefont{Kogut}}, \bibnamefont{and}
  \bibinfo{author}{\bibfnamefont{M.-P.} \bibnamefont{Lombardo}},
  \bibinfo{journal}{Nucl. Phys. Proc. Suppl.} \textbf{\bibinfo{volume}{60A}},
  \bibinfo{pages}{220} (\bibinfo{year}{1998}), \eprint{hep-lat/9705042}.

\bibitem[{\citenamefont{Alford et~al.}(1999)\citenamefont{Alford, Kapustin, and
  Wilczek}}]{Alford:1998sd}
\bibinfo{author}{\bibfnamefont{M.~G.} \bibnamefont{Alford}},
  \bibinfo{author}{\bibfnamefont{A.}~\bibnamefont{Kapustin}}, \bibnamefont{and}
  \bibinfo{author}{\bibfnamefont{F.}~\bibnamefont{Wilczek}},
  \bibinfo{journal}{Phys. Rev.} \textbf{\bibinfo{volume}{D59}},
  \bibinfo{pages}{054502} (\bibinfo{year}{1999}), \eprint{hep-lat/9807039}.

\bibitem[{\citenamefont{de~Forcrand and Laliena}(2000)}]{deForcrand:1999cy}
\bibinfo{author}{\bibfnamefont{P.}~\bibnamefont{de~Forcrand}} \bibnamefont{and}
  \bibinfo{author}{\bibfnamefont{V.}~\bibnamefont{Laliena}},
  \bibinfo{journal}{Phys. Rev.} \textbf{\bibinfo{volume}{D61}},
  \bibinfo{pages}{034502} (\bibinfo{year}{2000}), \eprint{hep-lat/9907004}.

\bibitem[{\citenamefont{Cox et~al.}(2000)\citenamefont{Cox, Gattringer,
  Holland, Scarlet, and Wiese}}]{Cox:1999nt}
\bibinfo{author}{\bibfnamefont{J.}~\bibnamefont{Cox}},
  \bibinfo{author}{\bibfnamefont{C.}~\bibnamefont{Gattringer}},
  \bibinfo{author}{\bibfnamefont{K.}~\bibnamefont{Holland}},
  \bibinfo{author}{\bibfnamefont{B.}~\bibnamefont{Scarlet}}, \bibnamefont{and}
  \bibinfo{author}{\bibfnamefont{U.~J.} \bibnamefont{Wiese}},
  \bibinfo{journal}{Nucl. Phys. Proc. Suppl.} \textbf{\bibinfo{volume}{83}},
  \bibinfo{pages}{777} (\bibinfo{year}{2000}), \eprint{hep-lat/9909119}.

\bibitem[{\citenamefont{Son and Stephanov}(2001{\natexlab{a}})}]{Son:2000xc}
\bibinfo{author}{\bibfnamefont{D.~T.} \bibnamefont{Son}} \bibnamefont{and}
  \bibinfo{author}{\bibfnamefont{M.~A.} \bibnamefont{Stephanov}},
  \bibinfo{journal}{Phys. Rev. Lett.} \textbf{\bibinfo{volume}{86}},
  \bibinfo{pages}{592} (\bibinfo{year}{2001}{\natexlab{a}}),
  \eprint{hep-ph/0005225}.

\bibitem[{\citenamefont{Son and Stephanov}(2001{\natexlab{b}})}]{Son:2000by}
\bibinfo{author}{\bibfnamefont{D.~T.} \bibnamefont{Son}} \bibnamefont{and}
  \bibinfo{author}{\bibfnamefont{M.~A.} \bibnamefont{Stephanov}},
  \bibinfo{journal}{Phys. Atom. Nucl.} \textbf{\bibinfo{volume}{64}},
  \bibinfo{pages}{834} (\bibinfo{year}{2001}{\natexlab{b}}),
  \eprint{hep-ph/0011365}.

\bibitem[{\citenamefont{Kogut and Sinclair}(2002{\natexlab{a}})}]{Kogut:2002tm}
\bibinfo{author}{\bibfnamefont{J.~B.} \bibnamefont{Kogut}} \bibnamefont{and}
  \bibinfo{author}{\bibfnamefont{D.~K.} \bibnamefont{Sinclair}},
  \bibinfo{journal}{Phys. Rev.} \textbf{\bibinfo{volume}{D66}},
  \bibinfo{pages}{014508} (\bibinfo{year}{2002}{\natexlab{a}}),
  \eprint{hep-lat/0201017}.

\bibitem[{\citenamefont{Kogut and Sinclair}(2002{\natexlab{b}})}]{Kogut:2002zg}
\bibinfo{author}{\bibfnamefont{J.~B.} \bibnamefont{Kogut}} \bibnamefont{and}
  \bibinfo{author}{\bibfnamefont{D.~K.} \bibnamefont{Sinclair}},
  \bibinfo{journal}{Phys. Rev.} \textbf{\bibinfo{volume}{D66}},
  \bibinfo{pages}{034505} (\bibinfo{year}{2002}{\natexlab{b}}),
  \eprint{hep-lat/0202028}.

\bibitem[{\citenamefont{Kogut et~al.}(2002)\citenamefont{Kogut, Toublan, and
  Sinclair}}]{Kogut:2002cm}
\bibinfo{author}{\bibfnamefont{J.~B.} \bibnamefont{Kogut}},
  \bibinfo{author}{\bibfnamefont{D.}~\bibnamefont{Toublan}}, \bibnamefont{and}
  \bibinfo{author}{\bibfnamefont{D.~K.} \bibnamefont{Sinclair}},
  \bibinfo{journal}{Nucl. Phys.} \textbf{\bibinfo{volume}{B642}},
  \bibinfo{pages}{181} (\bibinfo{year}{2002}), \eprint{hep-lat/0205019}.

\bibitem[{\citenamefont{Kogut et~al.}(2003)\citenamefont{Kogut, Toublan, and
  Sinclair}}]{Kogut:2003ju}
\bibinfo{author}{\bibfnamefont{J.~B.} \bibnamefont{Kogut}},
  \bibinfo{author}{\bibfnamefont{D.}~\bibnamefont{Toublan}}, \bibnamefont{and}
  \bibinfo{author}{\bibfnamefont{D.~K.} \bibnamefont{Sinclair}},
  \bibinfo{journal}{Phys. Rev.} \textbf{\bibinfo{volume}{D68}},
  \bibinfo{pages}{054507} (\bibinfo{year}{2003}), \eprint{hep-lat/0305003}.

\bibitem[{\citenamefont{Kogut and Sinclair}(2004)}]{Kogut:2004zg}
\bibinfo{author}{\bibfnamefont{J.~B.} \bibnamefont{Kogut}} \bibnamefont{and}
  \bibinfo{author}{\bibfnamefont{D.~K.} \bibnamefont{Sinclair}},
  \bibinfo{journal}{Phys. Rev.} \textbf{\bibinfo{volume}{D70}},
  \bibinfo{pages}{094501} (\bibinfo{year}{2004}), \eprint{hep-lat/0407027}.

\bibitem[{\citenamefont{de~Forcrand et~al.}(2007)\citenamefont{de~Forcrand,
  Stephanov, and Wenger}}]{deForcrand:2007uz}
\bibinfo{author}{\bibfnamefont{P.}~\bibnamefont{de~Forcrand}},
  \bibinfo{author}{\bibfnamefont{M.~A.} \bibnamefont{Stephanov}},
  \bibnamefont{and} \bibinfo{author}{\bibfnamefont{U.}~\bibnamefont{Wenger}},
  \bibinfo{journal}{PoS} \textbf{\bibinfo{volume}{LAT2007}},
  \bibinfo{pages}{237} (\bibinfo{year}{2007}), \eprint{0711.0023}.

\bibitem[{\citenamefont{Bernard et~al.}(1992)\citenamefont{Bernard, Kaiser,
  Kambor, and Meissner}}]{Bernard:1992qa}
\bibinfo{author}{\bibfnamefont{V.}~\bibnamefont{Bernard}},
  \bibinfo{author}{\bibfnamefont{N.}~\bibnamefont{Kaiser}},
  \bibinfo{author}{\bibfnamefont{J.}~\bibnamefont{Kambor}}, \bibnamefont{and}
  \bibinfo{author}{\bibfnamefont{U.~G.} \bibnamefont{Meissner}},
  \bibinfo{journal}{Nucl. Phys.} \textbf{\bibinfo{volume}{B388}},
  \bibinfo{pages}{315} (\bibinfo{year}{1992}).

\bibitem[{\citenamefont{Beane et~al.}(2005)\citenamefont{Beane, Bedaque,
  Parreno, and Savage}}]{Beane:2003yx}
\bibinfo{author}{\bibfnamefont{S.~R.} \bibnamefont{Beane}},
  \bibinfo{author}{\bibfnamefont{P.~F.} \bibnamefont{Bedaque}},
  \bibinfo{author}{\bibfnamefont{A.}~\bibnamefont{Parreno}}, \bibnamefont{and}
  \bibinfo{author}{\bibfnamefont{M.~J.} \bibnamefont{Savage}},
  \bibinfo{journal}{Nucl. Phys.} \textbf{\bibinfo{volume}{A747}},
  \bibinfo{pages}{55} (\bibinfo{year}{2005}), \eprint{nucl-th/0311027}.

\bibitem[{\citenamefont{Frink et~al.}(2002)\citenamefont{Frink, Kubis, and
  Meissner}}]{Frink:2002ht}
\bibinfo{author}{\bibfnamefont{M.}~\bibnamefont{Frink}},
  \bibinfo{author}{\bibfnamefont{B.}~\bibnamefont{Kubis}}, \bibnamefont{and}
  \bibinfo{author}{\bibfnamefont{U.-G.} \bibnamefont{Meissner}},
  \bibinfo{journal}{Eur. Phys. J.} \textbf{\bibinfo{volume}{C25}},
  \bibinfo{pages}{259} (\bibinfo{year}{2002}), \eprint{hep-ph/0203193}.

\bibitem[{\citenamefont{Tiburzi and Walker-Loud}(2008)}]{Tiburzi:2008bk}
\bibinfo{author}{\bibfnamefont{B.~C.} \bibnamefont{Tiburzi}} \bibnamefont{and}
  \bibinfo{author}{\bibfnamefont{A.}~\bibnamefont{Walker-Loud}},
  \bibinfo{journal}{Phys. Lett.} \textbf{\bibinfo{volume}{B669}},
  \bibinfo{pages}{246} (\bibinfo{year}{2008}), \eprint{0808.0482}.

\bibitem[{\citenamefont{Jiang and Tiburzi}(2009)}]{Jiang:2009sf}
\bibinfo{author}{\bibfnamefont{F.-J.} \bibnamefont{Jiang}} \bibnamefont{and}
  \bibinfo{author}{\bibfnamefont{B.~C.} \bibnamefont{Tiburzi}}
  (\bibinfo{year}{2009}), \eprint{0905.0857}.

\bibitem[{\citenamefont{Mai et~al.}(2009)\citenamefont{Mai, Bruns, Kubis, and
  Meissner}}]{Mai:2009ce}
\bibinfo{author}{\bibfnamefont{M.}~\bibnamefont{Mai}},
  \bibinfo{author}{\bibfnamefont{P.~C.} \bibnamefont{Bruns}},
  \bibinfo{author}{\bibfnamefont{B.}~\bibnamefont{Kubis}}, \bibnamefont{and}
  \bibinfo{author}{\bibfnamefont{U.-G.} \bibnamefont{Meissner}}
  (\bibinfo{year}{2009}), \eprint{0905.2810}.

\bibitem[{\citenamefont{Tiburzi}(2009)}]{Tiburzi:2009cf}
\bibinfo{author}{\bibfnamefont{B.~C.} \bibnamefont{Tiburzi}}
  (\bibinfo{year}{2009}), \eprint{0908.2582}.

\bibitem[{\citenamefont{Rentmeester et~al.}(2003)\citenamefont{Rentmeester,
  Timmermans, and de~Swart}}]{Rentmeester:2003mf}
\bibinfo{author}{\bibfnamefont{M.~C.~M.} \bibnamefont{Rentmeester}},
  \bibinfo{author}{\bibfnamefont{R.~G.~E.} \bibnamefont{Timmermans}},
  \bibnamefont{and} \bibinfo{author}{\bibfnamefont{J.~J.}
  \bibnamefont{de~Swart}}, \bibinfo{journal}{Phys. Rev.}
  \textbf{\bibinfo{volume}{C67}}, \bibinfo{pages}{044001}
  (\bibinfo{year}{2003}), \eprint{nucl-th/0302080}.

\bibitem[{\citenamefont{Fettes et~al.}(1998)\citenamefont{Fettes, Meissner, and
  Steininger}}]{Fettes:1998ud}
\bibinfo{author}{\bibfnamefont{N.}~\bibnamefont{Fettes}},
  \bibinfo{author}{\bibfnamefont{U.-G.} \bibnamefont{Meissner}},
  \bibnamefont{and}
  \bibinfo{author}{\bibfnamefont{S.}~\bibnamefont{Steininger}},
  \bibinfo{journal}{Nucl. Phys.} \textbf{\bibinfo{volume}{A640}},
  \bibinfo{pages}{199} (\bibinfo{year}{1998}), \eprint{hep-ph/9803266}.

\bibitem[{\citenamefont{Oller and Meissner}(2001)}]{Oller:2000fj}
\bibinfo{author}{\bibfnamefont{J.~A.} \bibnamefont{Oller}} \bibnamefont{and}
  \bibinfo{author}{\bibfnamefont{U.~G.} \bibnamefont{Meissner}},
  \bibinfo{journal}{Phys. Lett.} \textbf{\bibinfo{volume}{B500}},
  \bibinfo{pages}{263} (\bibinfo{year}{2001}), \eprint{hep-ph/0011146}.

\bibitem[{\citenamefont{Meissner et~al.}(2004)\citenamefont{Meissner, Raha, and
  Rusetsky}}]{Meissner:2004jr}
\bibinfo{author}{\bibfnamefont{U.~G.} \bibnamefont{Meissner}},
  \bibinfo{author}{\bibfnamefont{U.}~\bibnamefont{Raha}}, \bibnamefont{and}
  \bibinfo{author}{\bibfnamefont{A.}~\bibnamefont{Rusetsky}},
  \bibinfo{journal}{Eur. Phys. J.} \textbf{\bibinfo{volume}{C35}},
  \bibinfo{pages}{349} (\bibinfo{year}{2004}), \eprint{hep-ph/0402261}.

\bibitem[{\citenamefont{Borasoy et~al.}(2005)\citenamefont{Borasoy, Nissler,
  and Weise}}]{Borasoy:2005ie}
\bibinfo{author}{\bibfnamefont{B.}~\bibnamefont{Borasoy}},
  \bibinfo{author}{\bibfnamefont{R.}~\bibnamefont{Nissler}}, \bibnamefont{and}
  \bibinfo{author}{\bibfnamefont{W.}~\bibnamefont{Weise}},
  \bibinfo{journal}{Eur. Phys. J.} \textbf{\bibinfo{volume}{A25}},
  \bibinfo{pages}{79} (\bibinfo{year}{2005}), \eprint{hep-ph/0505239}.

\bibitem[{\citenamefont{Oller et~al.}(2005)\citenamefont{Oller, Prades, and
  Verbeni}}]{Oller:2005ig}
\bibinfo{author}{\bibfnamefont{J.~A.} \bibnamefont{Oller}},
  \bibinfo{author}{\bibfnamefont{J.}~\bibnamefont{Prades}}, \bibnamefont{and}
  \bibinfo{author}{\bibfnamefont{M.}~\bibnamefont{Verbeni}},
  \bibinfo{journal}{Phys. Rev. Lett.} \textbf{\bibinfo{volume}{95}},
  \bibinfo{pages}{172502} (\bibinfo{year}{2005}), \eprint{hep-ph/0508081}.

\bibitem[{\citenamefont{Gasser et~al.}(1991)\citenamefont{Gasser, Leutwyler,
  and Sainio}}]{Gasser:1990ce}
\bibinfo{author}{\bibfnamefont{J.}~\bibnamefont{Gasser}},
  \bibinfo{author}{\bibfnamefont{H.}~\bibnamefont{Leutwyler}},
  \bibnamefont{and} \bibinfo{author}{\bibfnamefont{M.~E.}
  \bibnamefont{Sainio}}, \bibinfo{journal}{Phys. Lett.}
  \textbf{\bibinfo{volume}{B253}}, \bibinfo{pages}{252} (\bibinfo{year}{1991}).

\bibitem[{\citenamefont{Buettiker and Meissner}(2000)}]{Buettiker:1999ap}
\bibinfo{author}{\bibfnamefont{P.}~\bibnamefont{Buettiker}} \bibnamefont{and}
  \bibinfo{author}{\bibfnamefont{U.-G.} \bibnamefont{Meissner}},
  \bibinfo{journal}{Nucl. Phys.} \textbf{\bibinfo{volume}{A668}},
  \bibinfo{pages}{97} (\bibinfo{year}{2000}), \eprint{hep-ph/9908247}.

\bibitem[{\citenamefont{Ohki et~al.}(2009)}]{Ohki:2009mt}
\bibinfo{author}{\bibfnamefont{H.}~\bibnamefont{Ohki}} \bibnamefont{et~al.}
  (\bibinfo{year}{2009}), \eprint{0910.3271}.

\bibitem[{\citenamefont{Walker-Loud et~al.}(2008)}]{WalkerLoud:2008bp}
\bibinfo{author}{\bibfnamefont{A.}~\bibnamefont{Walker-Loud}}
  \bibnamefont{et~al.} (\bibinfo{year}{2008}), \eprint{0806.4549}.

\bibitem[{\citenamefont{Walker-Loud}(2008)}]{WalkerLoud:2008pj}
\bibinfo{author}{\bibfnamefont{A.}~\bibnamefont{Walker-Loud}}
  (\bibinfo{year}{2008}), \eprint{0810.0663}.

\end{thebibliography}

\end{document}